\documentclass{aa}  
\usepackage{graphicx}
\usepackage{txfonts}
\usepackage{lipsum}
\usepackage{subcaption}      
\usepackage{lscape}          
\usepackage{placeins}        
\usepackage{xcolor}
\usepackage{natbib}
\usepackage[colorlinks=true, linkcolor=blue, citecolor=blue, urlcolor=blue]{hyperref}

\usepackage[normalem]{ulem}  
\begin{document}

   \title{The role of major mergers in triggering super-Eddington accretion}
   \author{Riccardo~Caleno\inst{1,2,3}\fnmsep\thanks{Corresponding author: riccardo.caleno@uniroma1.it}
        \and Tommaso~Zana\inst{1,2,3}
        \and Raffaella~Schneider\inst{1,2,3,4}
        \and Alessandro~Lupi\inst{5,6,7}
        \and Pedro~R.~Capelo\inst{8}
        \and Lucio~Mayer\inst{8}
        \and Alessandro~Trinca\inst{2,3,9}
        \and Rosa~Valiante\inst{2}
        \and Marta~Volonteri\inst{10}
        }

   \institute{Dipartimento di Fisica, ``Sapienza'' Universit\`{a} di Roma, Piazzale Aldo Moro 2, 00185 Roma, Italy 
   \and INAF, Osservatorio Astronomico di Roma, Via Frascati 33, 00040 Monte Porzio Catone, Italy
   \and INFN, Sezione Roma1, Dipartimento di Fisica, ``Sapienza'' Universit\`{a} di Roma, Piazzale Aldo Moro 2, 00185, Roma, Italy
   \and Sapienza School for Advanced Studies, Viale Regina Elena 291, 00161 Roma, Italy
   \and Como Lake Center for Astrophysics, DiSAT, Universit\`{a} degli Studi dell’Insubria, Via Valleggio 11, IT-22100 Como, Italy
   \and INFN, Sezione di Milano-Bicocca, Piazza della Scienza 3, IT-20126 Milano, Italy
   \and INAF, Osservatorio di Astrofisica e Scienza dello Spazio di Bologna, Via Gobetti 93/3, IT-40129 Bologna, Italy
   \and Department of Astrophysics, University of Zurich, Winterthurerstrasse 190, CH-8057 Z{\"u}rich, Switzerland
   \and Institute for Astronomy, University of Edinburgh, Royal Observatory, Blackford Hill, Edinburgh EH9 3HJ, UK
   \and Sorbonne Universit\'{e}, CNRS, UMR 7095, Institut d’Astrophysique de Paris, 98 bis bd Arago, F-75014 Paris, France}

   \date{Received May 24, 2026 / Accepted September 9, 2026}
 
  \abstract
  {James Webb Space Telescope observations have opened a new era in the exploration of the high-redshift Universe, revealing black holes (BHs) with masses of several million solar masses already at $z>8$, challenging our understanding of their growth mechanisms. In this context, super-Eddington (SE) accretion has emerged as a promising solution and has been widely adopted in both numerical simulations and semi-analytical models.}
  {In this work, we investigate whether a major merger between two relatively low-mass haloes ($M_{\rm halo}\sim10^9\,\mathrm{M_\odot}$) at high redshifts can trigger episodes of sustained SE accretion, with particular focus on the role of BH feedback.}
  {We employed state-of-the-art high-resolution cosmological zoom-in simulations of a major merger at $z\sim11$. We explored different prescriptions for BH seeding and feedback, including physically motivated radiative and kinetic models (winds and jets) across the three main accretion regimes: advection-dominated accretion flows, radiatively efficient sub-Eddington accretion, and SE accretion.}
  {For the relatively low-mass haloes studied here, our feedback prescription efficiently suppresses gas accretion, preventing substantial BH growth. We find that, although the merger drives gas inflows towards the central regions, this is not sufficient to trigger sustained SE accretion. Post-merger SE accretion episodes are observed only when BH feedback is entirely switched off. Amongst the feedback channels considered, kinetic feedback is the primary mechanism regulating BH growth. Moreover, the only significant SE accretion episodes occur immediately after BH seeding, while the merger itself does not produce a substantial enhancement of the accretion rate.  }
   {}


   \maketitle
   \nolinenumbers

\section{Introduction}

The discovery of supermassive black holes (SMBHs) by the James Webb Space Telescope (JWST) at very high redshift has resulted in challenges to previous models of BH formation and accretion. Observations of BHs with masses $M_{\rm BH}$ exceeding $10^6$~M$_\odot$ already at redshift $z \geq 8$ \citep[e.g.][]{kocevski2023, Kokorev2023, Harikane2023,Barro2024,Maiolino2024bhs,Maiolino2024z11, Tripodi2025, Taylor2025}, only a few hundred million years after the Big Bang, further challenge our understanding of early BH seeding and accretion mechanisms. 

James Webb Space Telescope observations have also revealed unusual properties of the host galaxies of these SMBHs. In particular, their BHs appear overmassive relative to the total stellar mass, $M_{*}$, of their host galaxies \citep[e.g.][]{Ubler2023, Bogdan2024, Brooks2026}, especially compared to the well-known local $M_{\rm BH}$--$M_*$ relation \citep[][]{Kormendy&Ho2013,Reines&Volonteri2015}. This discrepancy could be driven by selection effects \citep[see e.g.][]{Li2025} or even biases in how these masses have been measured \citep[e.g.][]{Lupi2024b, Trinca2026}; nonetheless, in a recent work \citep[][]{Juodzbalis2025}, the BH mass estimate for one of the most extreme objects has been measured dynamically. Therefore, even accounting for selection effects and the large intrinsic scatter \citep[][]{Ziparo2026}, these extreme systems still require a physical explanation.

The problem of overmassive BHs emerges when considering the classical picture of BH accretion theory. The So\l{}tan argument \citep[][]{Soltan_Argument} suggests that these BHs grow predominantly through gas accretion. In the framework of classical accretion theory, however, the accretion rate is typically limited. In spherically symmetric, steady state, and radiatively efficient conditions, this is reasonable since the two main regulators are gravity and radiation pressure acting over charged particles. This naturally imposes an upper limit on the luminosity of astrophysical objects, since if the radiative force overcomes the gravitational one, the system becomes unstable and the mass is driven away. This was first studied by Eddington \citep[][]{Eddington1916}. Assuming a given radiative efficiency, which we set to $1/16$ following \citet{Madau2014}, from the Eddington luminosity, $L_{\rm Edd}$, we can define the Eddington mass accretion rate as

\begin{equation}
    \dot{M}_{\mathrm{Edd}}=\frac{16\,L_{\rm Edd}}{c^2} =\frac{64\pi G M_{\mathrm{BH}}}{\kappa\, c}, 
\end{equation}

\noindent where $G$ is the gravitational constant, $\kappa$ is the opacity of the accreting material,\footnote{In general, and throughout this paper, we assume ionised hydrogen, such that $\kappa = \sigma_{\mathrm{T}}/m_{\mathrm{p}}$, where $\sigma_{\mathrm{T}}$ is the \citet{Thomson_1906} cross-section and $m_{\mathrm{p}}$ is the proton mass.} and $c$ is the speed of light in a vacuum.

Different proposals have been addressed to solve this problem, like primordial BHs \citep[PBHs; e.g.][]{Dayal2024,Ziparo2025,Prole2025,Zhang2026}, heavy seeds with masses $M_{\rm seed}\sim 10^{5-6}$~M$_\odot$ \citep[e.g.][]{Regan2009, Hosokawa2012, Mayer2015,Chon2016, Chon2025}, and super-Eddington (SE) accretion. Primordial BHs, which were initially developed to address the dark matter (DM) problem, appear as a particularly fascinating idea  \citep[see][for a review of the topic]{Villanueva2021}. However, the PBH scenario is subject to strong observational constraints, including limits from microlensing \citep[][]{Niikura2019}, ultra-faint dwarf galaxies \citep[][]{Stegmann_et_al_2020}, cosmic microwave background distortions, and gravitational-wave observations \citep[see][for a recent review on the topic]{Carr2026}, which restrict their abundance to a subdominant fraction of the DM. The formation of heavy seeds, on the other hand, requires specific environmental conditions \citep[][]{Schauer2017, Lupi2021MassiveSeeds,Latif2022, Mayer_et_al_2024}; nonetheless, they would need sustained accretion at nearly the Eddington limit for several million years, and this appears unlikely, primarily due to stellar feedback \citep[e.g. ][]{Dubois2015, Habouzit2017, Angles-Alcazar2017}. Finally, SE accretion has recently attracted significant attention within the scientific community, given observations over the past decades of ultraluminous X-ray sources and active galactic nuclei \citep[AGNs;][]{Begelman2006, Bachetti2014,Du2018, Tortosa2023}.

Historically, the first work proposing possible SE accretion mediated by photon trapping was by \citet{Begelman1979}. Later, \citet{Abramowicz1988} incorporated this idea into an accretion disc model, developing the so-called slim disc. Building on the earlier geometrically thin, optically thick disc model by \cite{Shakura&Sunayaev1973}, the slim disc extends the standard thin disc framework and recovers it at moderately sub-Eddington accretion rates \citep[][]{Abramowicz2013}. The key idea is that, above certain luminosities, the disc becomes geometrically thick enough for advection to become important, and the accretion flow is no longer radiatively efficient. This process has been investigated on small scales through very-high-resolution magnetohydrodynamic simulations, including a general relativistic treatment of the disc \citep[e.g.][]{Sadowski&Narayan2016, Dai2018, Curd2019}. 

Notably, semi-analytical models that include SE accretion are able to reproduce both the high BH-to-host mass ratio and the number density of the observed BH population \citep[][]{Trinca2022, Schneider2023, Geris2026}. This scenario also explains several peculiar characteristics of JWST-detected sources, such as their relatively weak X-ray emission \citep[][]{Madau2024, Inayoshi2025, Madau2026}.

In recent years, SE accretion has been implemented in numerical simulations through the development of sub-grid prescriptions. Several notable studies have focused on isolated simulations, investigating galactic cores \citep[e.g.][]{ Lupi2016, Sassano2023, Massonneau2023, Kao2026, Zana2025}. Some of these simulations can achieve sub-parsec spatial resolution and mass resolutions high enough to resolve individual stars, but they lack the large-scale environment required to follow their evolution over long timescales.

On cosmological scales, a number of simulations have extended accretion and feedback prescriptions to the SE regime, for example \citet{Zhu2022},  \citet{Ni2022}, \citet{Bhowmick2022}, \citet{Rennehan2024}, \citet{Bhowmick2026}, \citet{Chon2026} and  \citet{Chaikin2026}. In these works, however, the SE regime is typically modelled through sub-grid prescriptions that primarily account for the radiative component of the feedback, without including the kinetic component. 
Simulations that also include physically motivated kinetic feedback processes, such as jets or winds in the SE regime, which directly impact the interstellar medium of the host galaxy, remain relatively limited. Notable examples include \citet{Regan2019}, \citet{Takeo2020}, \citet{Lupi2024a}, \citet{Quadri2025} and \citet{Husko2026}.

The aim of this work is to investigate the role of major galaxy mergers in triggering SE accretion episodes. In particular, \citet{Lupi2024a} showed that sustained SE accretion can occur in highly overdense environments, such as massive quasar hosts. Here, we explore whether similar accretion episodes can also be triggered in lower-mass haloes, which would provide a natural pathway to explain the high AGN number densities inferred from recent JWST observations.

Several semi-analytical models \citep[e.g.][]{Pezzuli2016, Trinca2022, Izquierdo-Villalba2024, Trinca2026b} adopt a scenario in which merger-induced hydrodynamical \citep[e.g.][]{Capelo_Dotti_2017,Blumenthal_Barnes_2018} and gravitational \citep[e.g.][]{Hopkins_Quataert_2010,Capelo_et_al_2015} torques trigger gas inflows towards the central regions of galaxies, leading to phases of enhanced accretion onto the SMBH. To investigate this problem, we adopted the same numerical framework described in \citet{Lupi2024a} and \citet{Quadri2025}. In this work, however, we additionally accounted for the self-consistent evolution of the magnetic flux in the disc, as discussed in Section~\ref{subsubsection:BH_kin_feedback}.

The paper is organised as follows. In Section~\ref{section: Simulation code}, we briefly describe the code used and the main physical prescriptions adopted. In Section~\ref{section: runs}, we describe the simulation runs we performed. In Section~\ref{section: Results}, we present the results of the different runs, and finally, in Section~\ref{section: discussion&conclusions}, we discuss our findings and summarise our conclusions.

\section{Simulation code}\label{section: Simulation code}
 
The simulations were performed with the code \textsc{gizm} \citep[][]{Hopkins2015}, which employs the same domain decomposition and $N$-body algorithms as its predecessor \textsc{gadget} \citep[][]{Springel2005}, choosing its meshless finite-mass mode. We adopted a modified version of the code described in \citet{Lupi2024a} and \citet{Quadri2025}, which includes updated prescriptions for star formation (SF), stellar feedback, and radiative transfer \citep[][]{Lupi&Bovino2020, Lupi2021} as well as a physically motivated model for the SE accretion regime. In Appendix~\ref{Appendix:baryonic_physics}, we provide the details of the adopted baryonic physics.

The prescriptions describing BH physics are presented in detail in \citet{Lupi2024a, Quadri2025}. Here, only the most relevant aspects and the modifications adopted for the present simulations are summarised. 

\subsection{Black hole seeding and dynamics}

Black holes were seeded (unless otherwise specified) adopting a friends-of-friends based approach, already implemented in \textsc{gizmo}. The algorithm identifies groups of DM particles on the fly, adopting a linking length of $d_{\rm FoF} = 0.05$ in units of the mean interparticle separation of the high-resolution region, as in \citet{Lupi_et_al2019}. Gas, stellar, and BH particles were then associated with the group of their nearest DM particle, allowing galaxy properties to be reconstructed during runtime.

In each group that satisfies a given stellar or gas threshold criterion (see Table~\ref{tab:legend}) and does not already host a BH, a BH is seeded by converting a particle of the corresponding type: a gas particle when a gas-based threshold is adopted or a stellar particle when a stellar-mass threshold is used. Also, in some runs the BH is placed at the potential minimum of the halo, while in others it is placed at the highest-density peak (see Section~\ref{subsection: Initial conditions} for more details).

We accounted for unresolved dynamical friction effects from stellar and DM particles \citep[][]{Tremmel2015}, and we employed a decoupled dynamical mass for the BH \citep[][]{Angles-Alcazar2017}. In this approach, the BH is initialised at a physical mass of $M_{\rm BH}=10^4$~M$_\odot$, while its dynamical mass is set to $M_{\rm dyn}=2\times10^5$~M$_\odot$, a value 100 times larger than the baryonic particle mass. This choice prevents spurious scattering with surrounding particles. Furthermore, this approach avoids the need for ad hoc BH pinning schemes that artificially reposition the BH at the local potential minimum. Two BH particles are assumed to merge instantaneously when the less massive BH enters the smoothing kernel of the more massive BH. Gravitational-wave recoil is not included; consequently, the merger remnant receives no recoil kick.

\subsection{Black hole accretion}\label{subsubsection:BH_accretion}

We modelled accretion using the Bondi-Hoyle-Lyttleton \citep[BHL;][]{Hoyle_and_Lyttleton_1939, Bondi_and_hoyle_1944, Bondi_1952} formula:

\begin{equation}
\dot{M}_{\rm BH}
= \frac{4\pi\,G^{2}\,M_{\rm BH}^{2}\,\rho_{\rm gas}}
{\left(v^{2}+c_{\rm s}^{2}\right)^{3/2}},
\label{eq:bhl}
\end{equation}

\noindent where $\rho_{\rm gas}$ is the density of gas around the BH, $v$ is the local velocity of the gas with respect to the BH, and $c_{\rm s}$ is the local speed of sound.

The accretion rate was computed by kernel-averaging the individual contributions from the gas particles within this kernel. After each time step $\Delta t$, the code computed the mass increment as $\Delta M_{\rm BH}=(1-\eta_{\rm rad})\dot{M}_{\rm BH}\Delta t$, where $\eta_{\rm rad}$ is the radiative efficiency, discussed in detail below.

The BH kernel is defined as the region enclosing its 96 nearest gas neighbours. Therefore, it does not have a fixed size, but can vary from the softening length (see Section~\ref{subsection: Initial conditions}) in high-density regions up to a maximum value of 2~kpc in very-low-density region.

\subsection{Black hole feedback}

The code implements a physically motivated prescription for three accretion regimes classified by the Eddington ratio, $\lambda = \dot{M}_{\rm BH}/\dot{M}_{\rm Edd}$. The three regimes are

\begin{itemize}

    \item  $\lambda < 1.25\times10^{-3}$: advection-dominated accretion flow (ADAF),
    
    \item $1.25\times10^{-3} \le \lambda < 1$: sub-Eddington radiatively efficient accretion,
    
    \item $\lambda \ge 1$: SE accretion,
    
\end{itemize}

\noindent with different feedback prescriptions adopted for each regime. These prescriptions are particularly relevant to investigate the interplay between the SMBH and the host galaxy in  cosmological zoom-in simulations. We describe in the following the comprehensive approach to feedback implementation we have chosen, which includes both radiative and kinetic feedback.

\subsubsection{Radiative feedback}\label{subsubsection:BH_rad_feedback}

The radiative feedback efficiency was computed following \citet{Madau2014}, who derived a fitting formula for solutions of relativistic slim-disc equations presented by \citet{Sadowki2009}:

\begin{equation}
    \eta_{\mathrm{rad}} = \frac{A(a)}{16\lambda}\left[\frac{0.985}{1/\lambda + B(a)}+\frac{0.015}{1/\lambda + C(a)}\right],
    \label{eq:radiative_efficiency}
\end{equation}

\noindent where $A(a)$, $B(a)$, and $C(a)$ are functions of the BH spin parameter, $a= cJ/(GM_{\rm BH}^2)$, where $J$ is the magnitude of the BH angular momentum. These functions assume the expressions

\begin{equation}
    \begin{aligned}
        A(a) &= (0.9663 - 0.9292\,a)^{-0.5639},\\
        B(a) &= (4.627  - 4.445\,a)^{-0.5524},\\
        C(a) &= (827.3  - 718.1\,a)^{-0.7060},
    \end{aligned}
    \label{eq:radiative_efficiency_coefficients}
\end{equation}

\noindent with the implicit assumption of a prograde disc \citep[see][]{Kao2026}.

In the ADAF regime, the accretion flow becomes geometrically thick and optically thin, resulting in a lower radiative efficiency. To account for this effect, we suppressed the radiative efficiency defined in Equation~\eqref{eq:radiative_efficiency} by a factor $(\lambda/\lambda_{\rm ADAF})^{2/3}$ in the ADAF regime, with $\lambda_{\rm ADAF}=1.25\times 10^{-3}$ \citep[see][]{Xie&Yuan2012}.

The BH spin was assigned a fixed value in each run and was assumed to be aligned with the angular momentum of the gas within the BH kernel. We assumed that the infalling gas aligns the BH spin with its angular momentum on timescales shorter than the simulation time-step. This assumption is not always valid, and it has been partially explored in \citet{Kao2026}, but we leave the interplay between spin evolution and BH accretion to future work. 

The BH emission spectrum was modelled as the combination of a thermal black-body component and an X-ray corona, with a bolometric luminosity defined as

\begin{equation}
L_{\rm BH} = \eta_{\rm rad}\,\dot{M}_{\rm BH} c^{2}.
\end{equation}
The fraction of the bolometric luminosity emitted in the X-ray corona was determined using the relation from \citet{Duras2020}:

\begin{equation}
f^{-1}_{2-10\,\rm keV} \equiv \frac{L_{\rm BH}}{L_{2-10\,\rm keV}}
= 12.76 \left[1+\left(\frac{\log_{10}(L_{\rm BH}/{\rm L}_\odot)}{12.15}\right)^{18.78}\right].
\end{equation}
From this relation, the fraction of soft X-rays in the $0.2$--$2$~keV band was derived by assuming a coronal power-law spectrum with slope $-1.7$ \citep[][]{Regan2019}. The remaining fraction of the luminosity was attributed to the thermal (black-body) component and was computed as

\begin{equation}
f_{\rm BB} = 1 - f_{0.2-2\,\rm keV} - f_{2-10\,\rm keV}.
\end{equation}

Radiation transport was modelled using an on-the-fly M1 closure scheme \citep[][]{Levermore1984, Hopkins2020}, evolving the radiation energy density and flux adopting a reduced speed of light $c_{\rm red}=1000 \, \rm km \, s^{-1}$. Radiation was discretised into photon groups, each characterised by an energy density and flux, and was injected into gas elements inside the BH kernel.

The coupling between BH radiation and the surrounding gas occurred through photoionisation, photoheating, and radiation pressure. Photoionisation rates and photoheating were computed using \textsc{krome} \citep[][]{KROME}, which consistently accounts for the impact of the local radiation field on the ionisation balance and temperature. The momentum transfer from radiation to the gas was instead computed from the radiation flux, contributing to the acceleration of the gas, whereby gas elements with mass $m_{\rm gas}$ received a kick $m_{\rm gas} \Delta \mathbf{v} \propto (\Delta E_{\gamma}/c)$, where $\Delta E_{\gamma}$ is the photon energy injected \citep[][]{Lupi2020}.

\subsubsection{Kinetic feedback}\label{subsubsection:BH_kin_feedback}

The prescription for the kinetic component of the BH feedback follows the model presented in \citet{Lupi2024a}, with a minor modification introduced to improve the treatment of the magnetic flux regulating the jet power. The key parameters controlling the jet efficiency are the magnetic flux in the disc, $\phi$, and the critical value corresponding to a magnetically arrested disc (MAD), $\phi_{\rm MAD}$ \citep[][]{Narayan2003}. Instead of assuming a fixed value for the magnetic flux, we adopted the fitting formula for $\phi$ derived by \citet{Ricarte2023}\footnote{In this equation, for consistency with the reference paper, $f_{\rm Edd}$ denotes the Eddington fraction computed assuming the radiative efficiency of a thin disc, i.e. $\eta_{\rm rad}=1-\left(1-\frac{2}{3r_{\rm ms}}\right)^{1/2}$, where $r_{\rm ms}$ is the radius of the marginally stable orbit for a \citet{Kerr_1963} BH.},

\begin{equation}
\phi(a, f_{\rm Edd}) =
\phi_{\rm MAD}(a)\,
\frac{\left( f_{\rm Edd}/f_{\rm c} \right)^{\alpha}}
{1 + \left( f_{\rm Edd}/f_{\rm c} \right)^{\alpha}},
\label{eq.:magnetic flux}
\end{equation}

\noindent where $\alpha=1.29$ and $f_{\rm c}=1.88$. In Equation~\eqref{eq.:magnetic flux}, for $f_{\rm Edd}\gg f_{\rm c}$, we obtain $\phi\rightarrow\phi_{\rm MAD}$, while for $f_{\rm Edd}\ll f_{\rm c}$, $\phi\rightarrow0$ connecting to the thin disc solution. For the value of $\phi_{\rm MAD}$, we adopted a third-order polynomial fitting formula based on \citet{Narayan2022}, which depends only on the BH spin:

\begin{equation}
\phi_{\rm MAD}(a) =
52.6 + 34\,a - 14.9\,a^2 - 20.2\,a^3.
\label{eq.:mad}
\end{equation}

Given the spin and magnetic flux, we computed the electromagnetic jet efficiency, with the inclusion of higher-order correction factors \citep[][]{Tchekhovskoy2015, Pan&Yu2015},

\begin{equation}
\eta_{\rm jet} =
\frac{k}{4\pi}\,
\phi^{2}\Omega_{\rm H}^{2}
\left(1 + 1.38\,\Omega_{\rm H}^{2} - 9.2\,\Omega_{\rm H}^{4}\right),
\label{eq.:kin eff}
\end{equation}

\noindent where $\Omega_{\rm H}$ is the angular velocity computed at the BH horizon, assuming a \citet{Kerr_1963} spacetime, 

\begin{equation}
\Omega_{\rm H} \equiv \frac{|a|}{2 r_{\rm H}}
= \frac{|a|}{2\left(1 + \sqrt{1 - a^{2}}\right)},
\label{eq.:ang_mom}
\end{equation}

\noindent and $k$ is a constant that depends on the initial geometry, and we assumed it is equal to $k = 0.05$, as in \citet{Ricarte2023}. The jets were launched along the BH spin axis in a region delimited by a cylinder whose base is assumed equal to the BH kernel. To compute the mass loading factor of jets, $\beta_{\rm jet}\equiv\dot{M}_{\rm jet}/\dot{M}_{\rm BH}$, we assumed a jet velocity of $v_{\rm jet}=0.1c$, and from energy conservation, the resulting mass loading factor is given by

\begin{equation}
    \beta_{\rm jet} = \frac{\dot{M}_{\rm jet}}{\dot{M}_{\rm BH}} = 2 \,\eta_{\rm jet} \left( \frac{c}{v_{\rm jet}} \right)^2.
    \label{eq.:mass loading factor}
\end{equation}

The kinetic component of the feedback was modelled as jets for the ADAF and SE regimes, while for the sub-Eddington radiatively efficient regime ($1.25\times10^{-3} < \lambda < 1$) the disc is well described by the thin disc solution \citep[][]{Shakura&Sunayaev1973}. In this regime, the MBH kinetic feedback was assumed to be released in the form of bipolar line-driven winds. The mass loading of the winds was determined from momentum conservation during the matter-radiation interaction \citep[][]{Choi2012, Angles-Alcazar2017}, and $v_{\rm jet}=v_{\rm w}=0.1c$:

\begin{equation}
\beta_{\rm w} = 
\frac{L_{\rm BH}}{\dot{M}_{\rm BH} v_{\rm w} c}
= \eta_{\rm rad}\frac{c}{v_{\rm w}},
\end{equation}
The corresponding energy coupling efficiency of the winds was estimated as

\begin{equation}
\epsilon_{\rm w} =
\frac{\dot{M}_{\rm w} v_{\rm w}^{2}}{2L_{\rm BH}} =
\frac{\dot{M}_{\rm w} v_{\rm w}^{2}}{2\eta_{\rm rad}\dot{M}_{\rm BH}c^{2}} =
\frac{\beta_{\rm w} v_{\rm w}^{2}}{2\eta_{\rm rad} c^{2}} =
\frac{v_{\rm w}}{2c}.
\end{equation}

For the sub-Eddington and SE regime, the coupling efficiency of the winds was then compared with that associated with the jet component. Whenever the wind efficiency was higher than that of the jets, feedback was instead injected through the wind channel, following the prescription described above. This was done to avoid vanishing feedback at very low BH spin accounting for the fact that SE accretion episodes are also expected to launch radiative outflows even when the magnetisation is low. Also in this case the winds are launched along the angular momentum of the BH, differently from the previous version, which assumed an opening angle of 45 degrees \citep[][]{Sala2021}. Given the resolution of our simulation, we do not expect that the opening of the cone significantly changes the results. 

Numerically, after each accretion episode, gas particles within the BH kernel are stochastically selected for accretion according to a kernel-weighted probability. Selected gas particles simultaneously contribute to both accretion and kinetic feedback. In particular, only a fraction $f_{\rm acc} = 1/(1+\beta_{\rm w/jet})$ of their mass is accreted onto the BH, while the remaining fraction $(1 - f_{\rm acc})$ is retained by the particle and receives a velocity kick, depending on the feedback mode. Radiative feedback, instead, is coupled to all the particles contained in the BH kernel.

\begin{figure*}
    \centering
    \includegraphics[width=17cm]{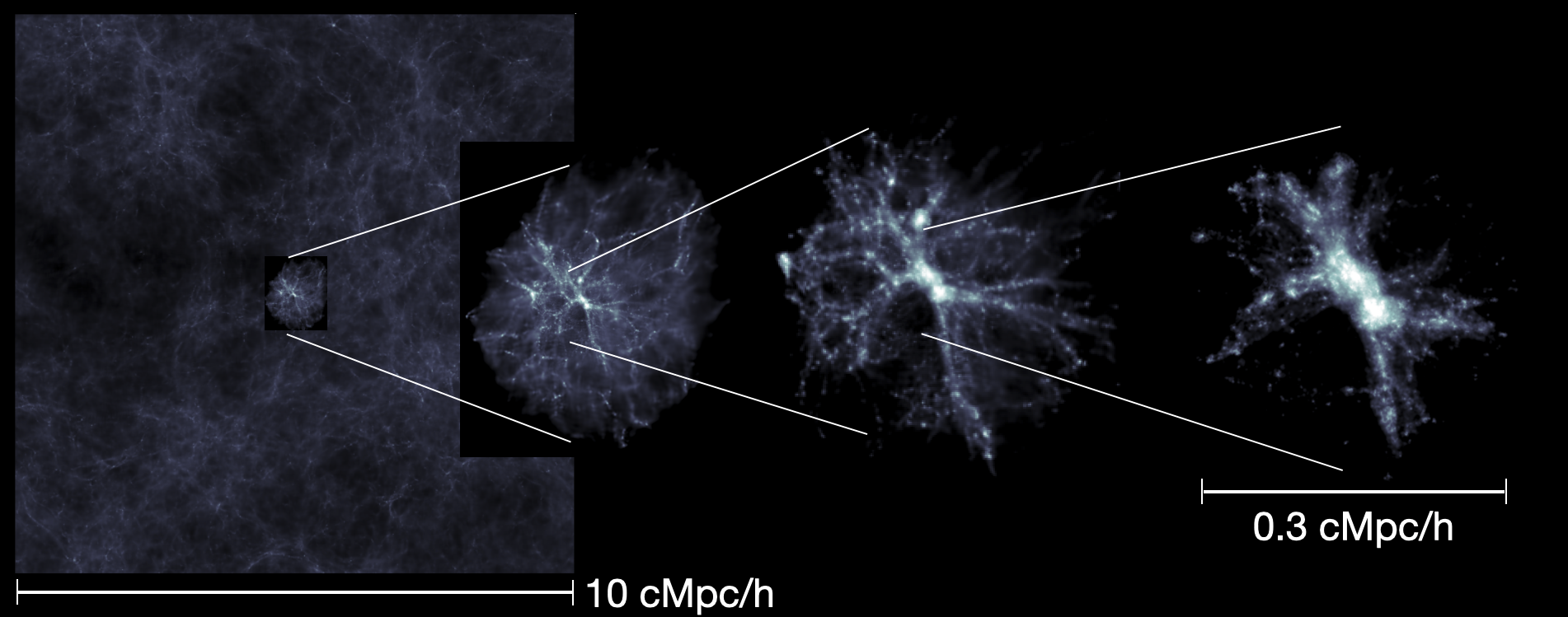}
    \caption{Zoom-in DM density snapshots at the moment of the merger ($z=11.03$). Every zoomed region represents a different refinement level. Each level has a mass resolution $2^{-3}$ times the previous one,  starting with a DM mass resolution of $\sim$$5\times10^6$~M$_\odot$, reaching in the last level a resolution of $\sim$$\rm10^4$~M$_\odot$.} 
    \label{fig:zoom-in}
\end{figure*}

\section{Runs}\label{section: runs}

\begin{table*}[]
    \centering
    \begin{tabular}{|c|c|c|c|c|c|c|c|}
     \hline
       Name & Seeding method & $z_{\rm end}$ & Seeding condition & BH spins & Rad. feedback & Kin. feedback & Seed location\\
      \hline
      Run1 & $M_\star$ threshold & 10 & $M_\star\geq10^6$~M$_\odot$ & 0.7 & on & on & potential minimum\\
      Run2 & $M_{\rm gas}$ threshold & 10 & $M_{\rm gas}\geq10^6$~M$_\odot$ & 0.7 & on & on & potential minimum\\
      Run3 & ad hoc seed & 10 & $\rm SFR>0$ & 0.7 & on & on & density peak\\
      Run4 & ad hoc seed noFB & 10 & $\rm SFR>0$ & 0 & off & off & density peak\\
      Run5 & $M_\star$ threshold & 8 & $M_\star\geq10^6$~M$_\odot$ & 0.7 & on & on & density peak\\
      Run6 & $M_\star$ threshold & 8 & $M_\star\geq10^6$~M$_\odot$ & 0.2 & on & on & density peak\\
      Run7 & $M_\star$ threshold & 8 & $M_\star\geq10^6$~M$_\odot$ & 0.7 & on & off & density peak \\
     \hline
\end{tabular}
    \caption{Summary of the different runs, with different final redshifts, seeding prescriptions, BH spins, BH feedback, and seed locations.}
    \label{tab:legend}
\end{table*}

\subsection{Initial conditions}\label{subsection: Initial conditions}

We performed a total of seven different runs, summarised in Table~\ref{tab:legend}. We started from a DM-only parent simulation with a box length of 10~cMpc~$h^{-1}$ at $z_{\rm ic}=100$. The initial conditions were generated using \textsc{music} \citep[][]{MUSIC}, adopting the cosmological parameters from \citet{Planck2016}: $\Omega_{\Lambda}=0.6911$, $\Omega_{\mathrm{m}}=0.3089$, $\Omega_{\mathrm{b}}=0.0489$, $h=0.6774$, $n_{\rm s}=0.9667$, and $\sigma_8=0.8159$. From this simulation, we selected a major merger event between two haloes with $m_1= 2 \times10^9$~M$_\odot$ and $m_2= 8\times 10^8$~M$_\odot$ ($m_2/m_1\approx0.4$) occurring at $z = 11$. The resulting post-merger DM halo mass is $\approx 3\times 10^9$~M$_\odot$, much smaller than the halo simulated in the study of \citet{Lupi2024a}, who were targeting the host of a quasar ($3\times 10^{12}$~M$_\odot$ at $z=6$). We then recursively refined a Lagrangian region extending up to $2.5$ virial radii\footnote{We define the virial radius as $r_{\rm vir}\equiv r_{200\rm c}$, namely the radius enclosing a mean density equal to 200 times the critical density of the Universe at the corresponding redshift.} of the target halo at a chosen redshift $z_{\rm end}$, corresponding to the final time of each simulation, following the procedure described in \citet{Fiacconi_2017}. All particles located within $2.5$ virial radii of the target halo at $z_{\rm end}$ were traced back to their positions at the beginning of the simulation. \textsc{music} then computed the minimum ellipsoid enclosing these particles at $z_{\rm ic}$, which defined the high-resolution region. This procedure was iteratively repeated, producing a sequence of consecutive ellipsoids with progressively decreasing resolution levels (see Figure~\ref{fig:zoom-in}), and this was done to avoid contamination by low-resolution DM particles within the virial radius.

Since our goal is to investigate the impact of the merger on BH accretion, we set the final redshift of the first four runs to $z_{\rm end}=10$. This choice provides approximately 50~Myr after the merger, allowing the gas and BHs sufficient time to settle into the potential well of the resulting halo. After that, in order to study possible late time merger effects on the gas inflows and impact on BH accretion, we performed three additional runs, with final time $z_{\rm end}=8$ (see Table~\ref{tab:legend})\footnote{The runs evolved to $z_{\rm end}=10$ required, on average, approximately $3\text{--}4\times10^{4}$ CPU hours per run, corresponding to about 250–350 hours of wall-clock time on 112 cores. The runs evolved to $z_{\rm end}=8$ required approximately $1\text{--}1.2\times10^{5}$ CPU hours per run, corresponding to about 300–350 hours of wall-clock time on 336 cores.}. The identification of the haloes and their merger history was performed using the \textsc{amiga halo finder} \citep[][]{Gill_et_al_2004, AHF2009}. 

The gravitational softening of the simulation, at the highest refinement level, is 10 and 120~pc for stars and DM, and 1~pc for gas and BHs. All these quantities are given in comoving units and are kept constant throughout the simulation. The mass resolution at the same level is $\sim 10^3$~M$_\odot$ for baryons and $\sim 10^4$~M$_\odot$ for DM.

\subsection{Setup of the simulations}

In these runs, different seeding mechanisms and AGN feedback prescriptions were systematically explored. Each run is described below, while the main characteristics of the simulations are summarised in Table~\ref{tab:legend}. Unless otherwise stated, the BHs were assumed\footnote{We adopted this value because it yields the commonly assumed radiative efficiency in the standard sub-Eddington, radiatively efficient accretion regime: $\eta_{\rm rad}\sim 0.1$.} to form with a constant spin of $a=0.7$.

\begin{enumerate}

    \item $M_\star$ threshold (Run1, $z_{\rm end}=10$): We seeded BHs in the potential minimum of haloes once their stellar mass reached $M_\star = 10^6$~M$_\odot$. With this prescription, two BHs formed, both seeded at $z\approx15$ in the haloes that later underwent the major merger at $z=11$.

    \item $M_{\rm gas}$ threshold (Run2, $z_{\rm end}=10$): In this case, we seeded BHs in the halo potential minima when the total gas mass exceeded $M_{\rm gas}=10^6$~M$_\odot$. Since this condition is easier to satisfy, this run produced several tens of BHs, significantly more than in Run1.

    \item Ad hoc seed (Run3, $z_{\rm end}=10$): In this run, we adopted a different seeding strategy. Starting from the configuration of the $M_\star$ threshold run, we seeded two BHs in the same haloes, but at the highest gas density peaks satisfying a SF rate ${\rm SFR}>0$. This occurred at $z\sim22$, and the corresponding densities were $\sim$$10^{3}~{\rm cm^{-3}}$ and $\sim$$10^{2}~{\rm cm^{-3}}$ for the two peaks\footnote{These number densities were obtained from the gas mass density from $n=\rho/(\mu m_{\rm p})$, assuming ionised hydrogen, i.e. with a mean molecular weight $\mu\approx 0.59$.}. Due to this difference, in the higher-density peak we seeded the BH with the standard physical and dynamical mass, while for the second BH we adopted a lower physical mass of $M_{\rm BH}=10^3$~M$_\odot$, keeping the same dynamical mass.

    \item Ad hoc seed without feedback (Run4, $z_{\rm end}=10$): This run followed the same setup as Run3 but with BH feedback completely disabled, both radiative and kinetic, in this case the BH spin is irrelevant and was therefore set to $a=0$. The purpose of this simulation was to study BH accretion in an idealised environment and to estimate the upper limit of the accretion rates achievable in the absence of feedback.

    \item $M_\star$ threshold and density peak (Run5, $z_{\rm end}=8$): Similarly to Run1, we seeded BHs once the stellar mass threshold $M_\star = 10^6$~M$_\odot$ was reached, but in this case at the density peaks of the haloes rather than at the potential minimum, since this configuration was found to produce higher accretion rates.

    \item Reduced spin (Run6, $z_{\rm end}=8$): This run adopted the same setup as Run5 but with a lower BH spin, $a=0.2$, in order to explore a regime with reduced jet power and therefore weaker kinetic feedback (see Equations~\ref{eq.:kin eff} and \ref{eq.:ang_mom}).

    \item No kinetic feedback (Run7, $z_{\rm end}=8$): Finally, we performed a third simulation with the same seeding prescription as Run5, but neglecting the kinetic feedback component.
    
\end{enumerate}

\section{Results}\label{section: Results}

\subsection{Cosmological environment}

\begin{figure}
    \resizebox{\hsize}{!}{\includegraphics{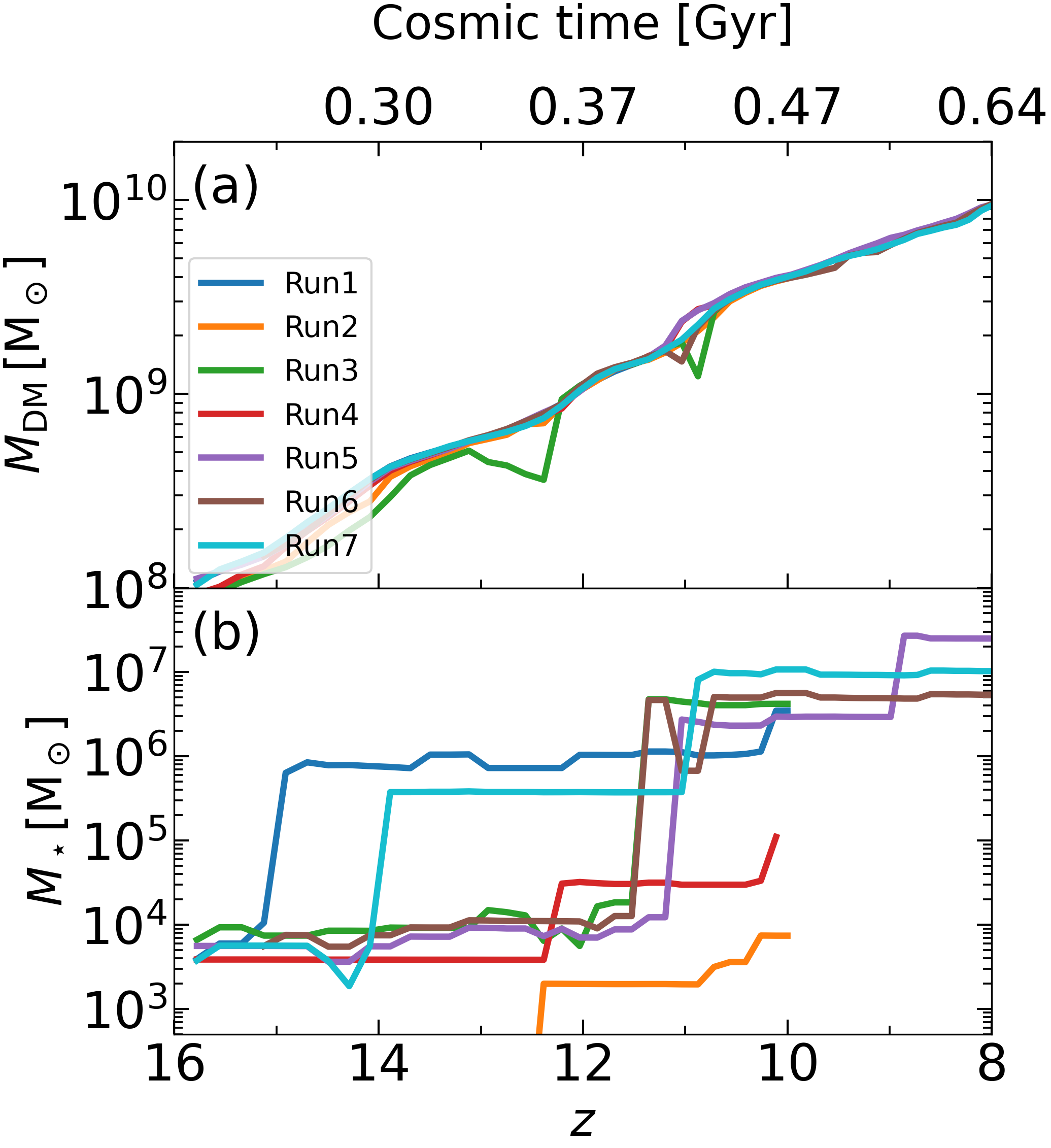}}
    \caption{Evolution of the DM mass (upper panel) and of the stellar mass (lower panel) of the main halo as a function of redshift for all seven runs. The halo centre and virial radius were determined with the \textsc{amiga halo finder}. The DM mass was computed within $r_{\rm vir}$, 
while the stellar mass was measured within $0.3\,r_{\rm vir}$ after 
recentring on the stellar density peak associated with the central galaxy. The DM mass evolution is nearly identical across the simulations, whereas the stellar mass shows significant variation amongst the runs.}
    \label{fig:halo_properties}
\end{figure}

In Figure~\ref{fig:halo_properties}, we show the evolution of the main halo, i.e. the progenitor of the final post-merger system, in all simulations. With this definition, at $z \gtrsim 11$ we follow the more massive halo (with mass $m_1 = 2\times10^9$~M$_\odot$) that will later merge, while at $z \lesssim 11$ we follow the resulting post-merger halo. Unless otherwise stated, the halo mass is calculated within the virial radius, $r_{\rm vir}$, determined by the \textsc{amiga halo finder}. For the stellar component, we use AHF to identify the main halo and determine 
$r_{\rm vir}$, but we independently determine the centre of the stellar 
distribution. In particular, we recentre on the stellar density peak 
associated with the galaxy and compute the stellar mass within a 
sphere of radius $0.3\,r_{\mathrm{vir}}$ around this position \citep[e.g.][]{Zana2022}.

The upper panel shows the DM mass as a function of redshift, while the lower panel presents the stellar mass of the halo. At $z\sim15$, the main haloes have DM masses of $\sim10^{8}$~M$_\odot$, and in the runs evolved to $z=10$ they reach final DM masses of $\sim10^{10}$~M$_\odot$. To ensure that the same main halo was consistently tracked across all simulations, we adopted the following identification procedure. Using the merger tree of Run5 as a reference, at each snapshot we searched in the other simulations for haloes located within $0.3\,r_{\rm vir}$ of the reference main halo. Amongst these candidates, we selected the halo whose mass is closest to that of the reference halo.

The DM mass histories are very similar in all runs. Small differences appear around $z\approx11$, when the two haloes approach the merger. At this stage, the halo finder struggles to consistently identify the same halo across the different simulations because of the highly dynamical nature of the interaction. A small fluctuation is visible in Run3 at $z\simeq12.5$, likely reflecting the chaotic environment of the system, where baryonic processes can perturb the halo properties \citep[][]{Zana2022}.

The lower panel shows the evolution of the stellar mass of the central galaxy. In contrast to the DM component, significant differences appear amongst the runs. The step-like behaviour highlights the bursty nature of SF in this environment. At these early times, the stellar component can be 
offset from the centre of the DM halo during the highly dynamical assembly 
phase. Recentring the stellar-mass aperture on the stellar density peak 
allows us to consistently follow the central stellar component and avoids 
spurious variations caused by changes in this offset. We recover that the increase to $M_\star \gtrsim 3\times10^{6}$~M$_\odot$ around $z\approx11$ is a direct consequence of the merger: the gas inflows generated by the merger enhance the gas density, boosting the SFR, and consequently increasing the total stellar mass. A smaller but non-negligible fraction is contributed by stars formed ex situ in the secondary halo, which typically account for $\sim$$10$--$20\%$ of the post-merger stellar mass.

A particularly clear example of the run-to-run variation is provided by Runs~1 and~5 at $z\gtrsim11.5$, wherein their stellar masses differ by up to approximately two orders of magnitude. Although the two simulations target the same halo and adopt the same BH seeding prescription, initially small differences in the timing and spatial clustering of SF are non-linearly amplified by SN feedback. In Run~5, the SN energy injected at $z\sim17$ is comparable to the binding energy of the halo. This early feedback event heats and removes gas from the central region, suppressing subsequent SF. Run1, by contrast, experiences a weaker early injection of SN energy, allowing more gas to remain in the central region, where it continues to accumulate and form dense structures. This eventually produces a strong burst of SF around $z\sim15$. The large difference in stellar mass therefore arises from a feedback-driven divergence between the evolutionary histories of the two haloes. This effect is particularly strong because of the low masses and shallow gravitational potentials considered, which make the central gas reservoirs highly susceptible to individual or temporally clustered SN events.

The increase in stellar mass, however, strongly depends on the presence and number of massive BHs. Runs hosting a moderate number of BHs (at most two, one per progenitor halo) and including radiative feedback reach broadly similar stellar masses ($10^{6-7}$~M$_\odot$) by $z=10$ (see runs~1, 3, 5, 6, and 7 in Figure~\ref{fig:halo_properties}). In runs~1 and~7, the stellar mass is slightly higher before the merger compared to the other runs. This reflects the stochastic nature of SF (see Appendix~\ref{Appendix:baryonic_physics}), which can lead to a larger or smaller number of stars forming in different haloes. 

\begin{figure*}
    \centering
    \includegraphics[width=17cm]{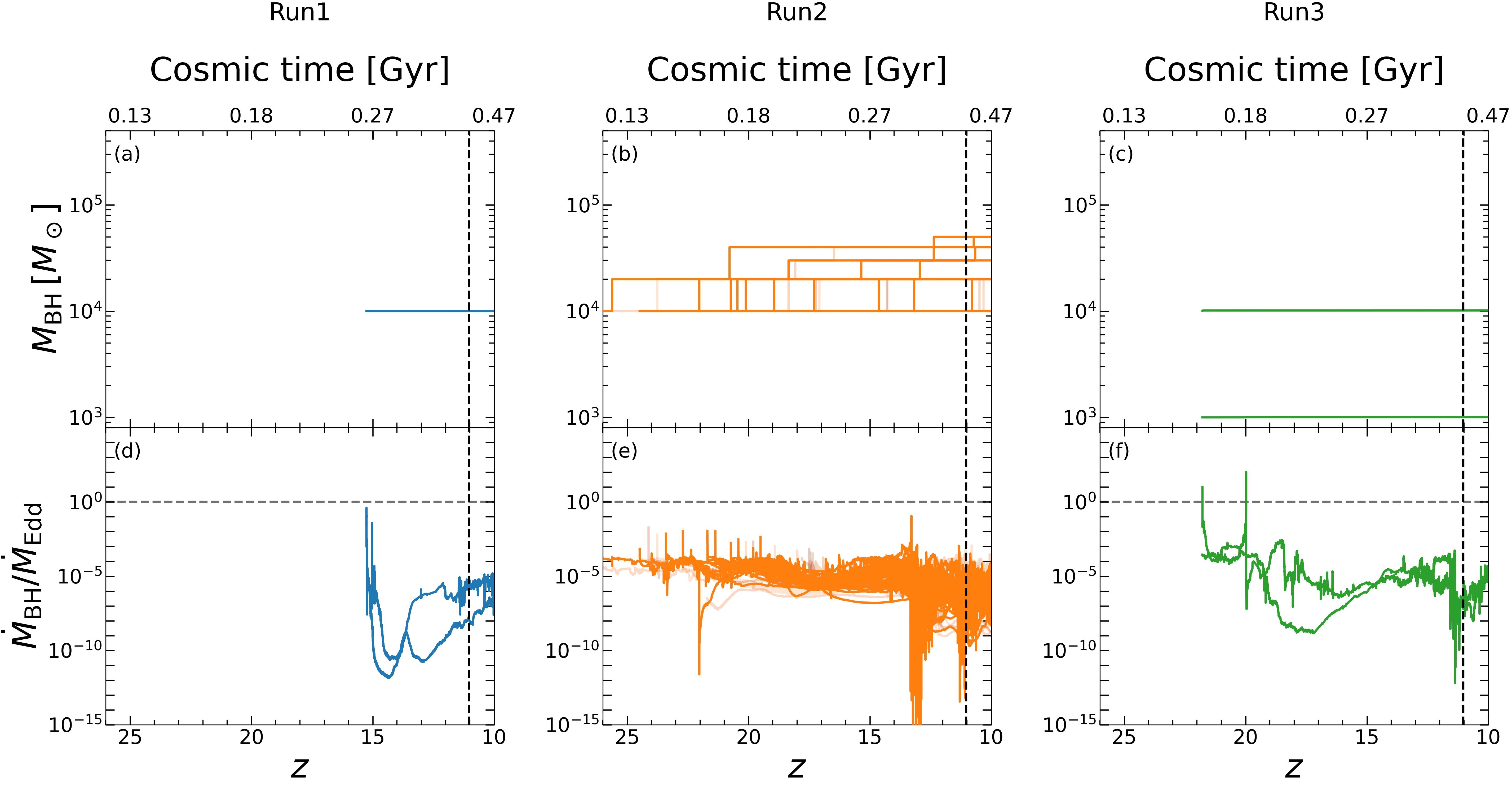}
    \caption{Comparison amongst three different seeding prescriptions: the simulation with $M_\star$ threshold (left-hand panels), the run with $M_{\rm gas}$ threshold (middle panels), and the ad hoc seed run (right-hand panels). In the upper plots we show the evolution of the BH mass, and in the lower panels we show the accretion rate in terms of the Eddington rate. The BHs located within the virial radius of the final main halo are highlighted with thicker, fully opaque curves, whereas the remaining BHs are shown with thinner and partially transparent lines. The colours have been specifically chosen to represent a specific run, as in Figure~\ref{fig:halo_properties}. In the lower panels, the dashed lines represent the Eddington rate ($\lambda = 1$). The vertical dashed line, represents the moment of the merger.}
    \label{fig:seeding_comparison}
\end{figure*}

Run2 shows a markedly different behaviour. The final stellar mass is considerably lower than in the other simulations. Two effects contribute to this outcome. First, BHs are seeded directly from gas particles located at the bottom of the halo potential well, regions that would otherwise efficiently form stars. Second, the large number of BHs formed in this run generates strong cumulative BH feedback, which suppresses the SFR. As a consequence, the stellar mass remains significantly lower than in the other runs (see Figure~\ref{fig:halo_properties}).

Run4 also results in a relatively low final stellar mass. This behaviour is linked to the adopted seeding strategy combined with the absence of feedback. BHs are seeded at high-density peaks before any stellar particles form, and in the absence of feedback the surrounding gas is preferentially accreted onto the BH rather than converted into stars. Gas that would otherwise form stars therefore contributes to BH growth, resulting in a smaller stellar component in the final halo. This effect is not observed in Run3, where kinetic feedback rapidly suppresses BH growth, nor in Run7, where BHs are seeded at later times, when SF has already started.

\subsection{Dependence on seeding prescription}\label{subsection: Seeding dependence} 

To assess the impact of the seeding prescription on the accretion onto BHs, we compare the three seeding mechanisms adopted in this work: the $M_\star$ threshold, the $M_{\rm gas}$ threshold, and the ad hoc seeding, respectively Run1, Run2, and Run3. The results are shown in Figure~\ref{fig:seeding_comparison}, wherein each column corresponds to one run. 

In Run1, BHs are seeded at $z\sim15$, and their subsequent accretion remains negligible throughout the simulation. In Run2, by contrast, several tens of BHs form, since the seeding condition is much easier to satisfy. However, none of them experiences an SE accretion phase. By the end of the simulation, some BHs reach masses of $\sim5\times10^{4}$~M$_\odot$, mainly through successive mergers, as clearly visible from panel (b) in Figure~\ref{fig:seeding_comparison}. 

A different behaviour is observed in Run3. In this case both BHs undergo initial episodes of SE accretion [panel (f) of Figure~\ref{fig:seeding_comparison}], triggered by the seeding in high-density regions. These episodes, however, are, quite brief - of the order of 1--10~kyr, see panel (d) of Figure~\ref{fig:all_runs} - and do not significantly affect the final BH masses. 

These results indicate that, although the seeding prescription can trigger initial SE accretion episodes, as in the ad hoc seeding case, this alone is not sufficient to sustain accretion over long timescales. As shown below, this is due to the strong BH feedback, in particular its kinetic component.

\subsection{Dependence on black hole feedback}\label{subsection: feedback_dependence}

\begin{figure*}
\sidecaption
  \includegraphics[width=12cm]{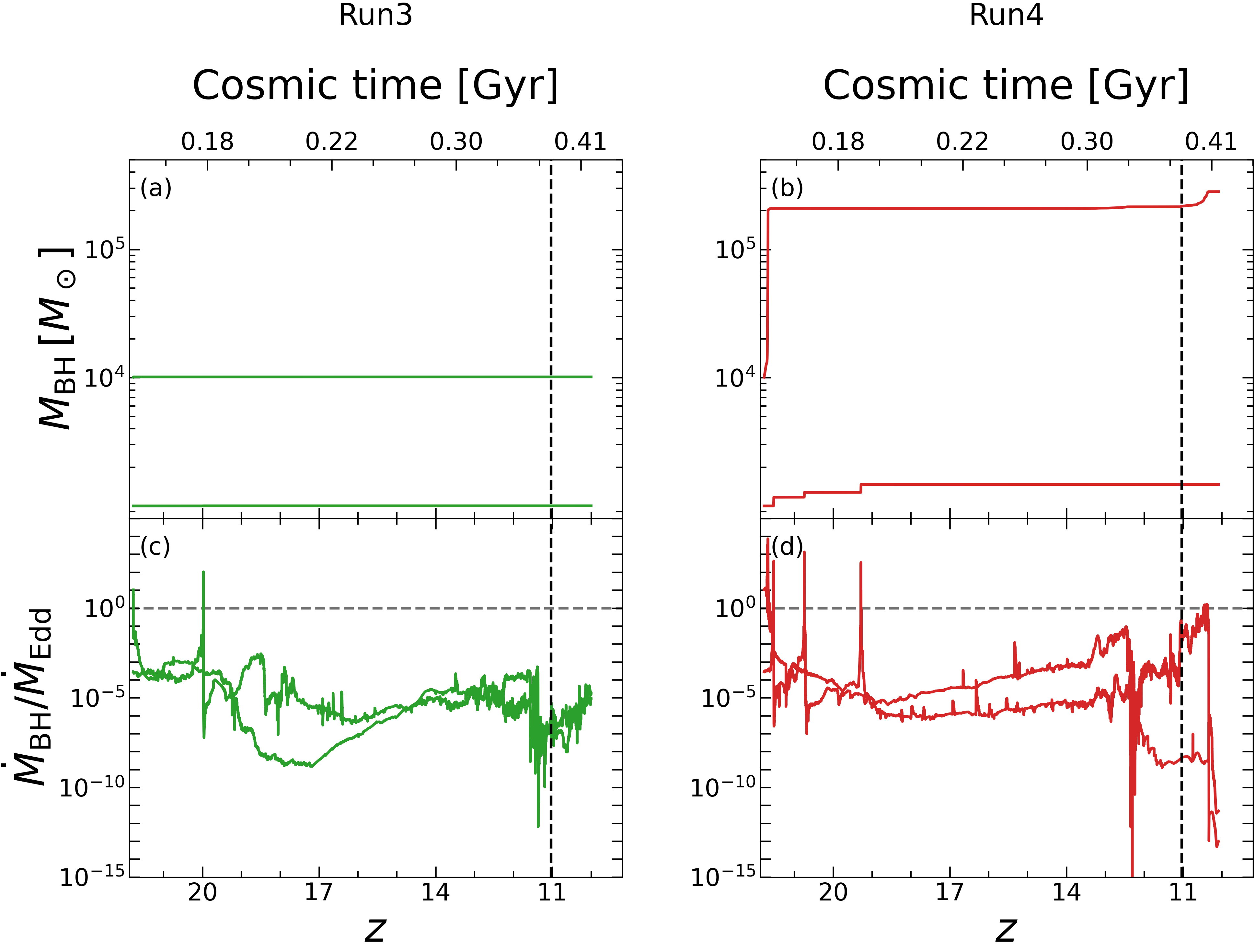}
  \caption{Comparison between ad hoc seed and ad hoc seed noFB runs. We show the evolution of BH masses and the BH accretion rates as in Figure~\ref{fig:seeding_comparison}. The effect on BH accretion is clear: In the run without feedback (on the right), the most massive BH grows through an initial burst to a few times $10^5$M$_\odot$, while in the run with kinetic and radiative feedback (on the left), it remains roughly at $10^4$M$_\odot$  until the end of the simulation.}
  \label{fig:feedback_comparison}
\end{figure*}

In order to isolate the role of BH feedback in our simulations, we performed a simulation in which both kinetic and radiative feedback were disabled. We adopted the same seeding conditions of Run3, allowing for a direct comparison between the two cases. The results are shown in Figure~\ref{fig:feedback_comparison}.

\begin{figure}
    \resizebox{\hsize}{!}{\includegraphics{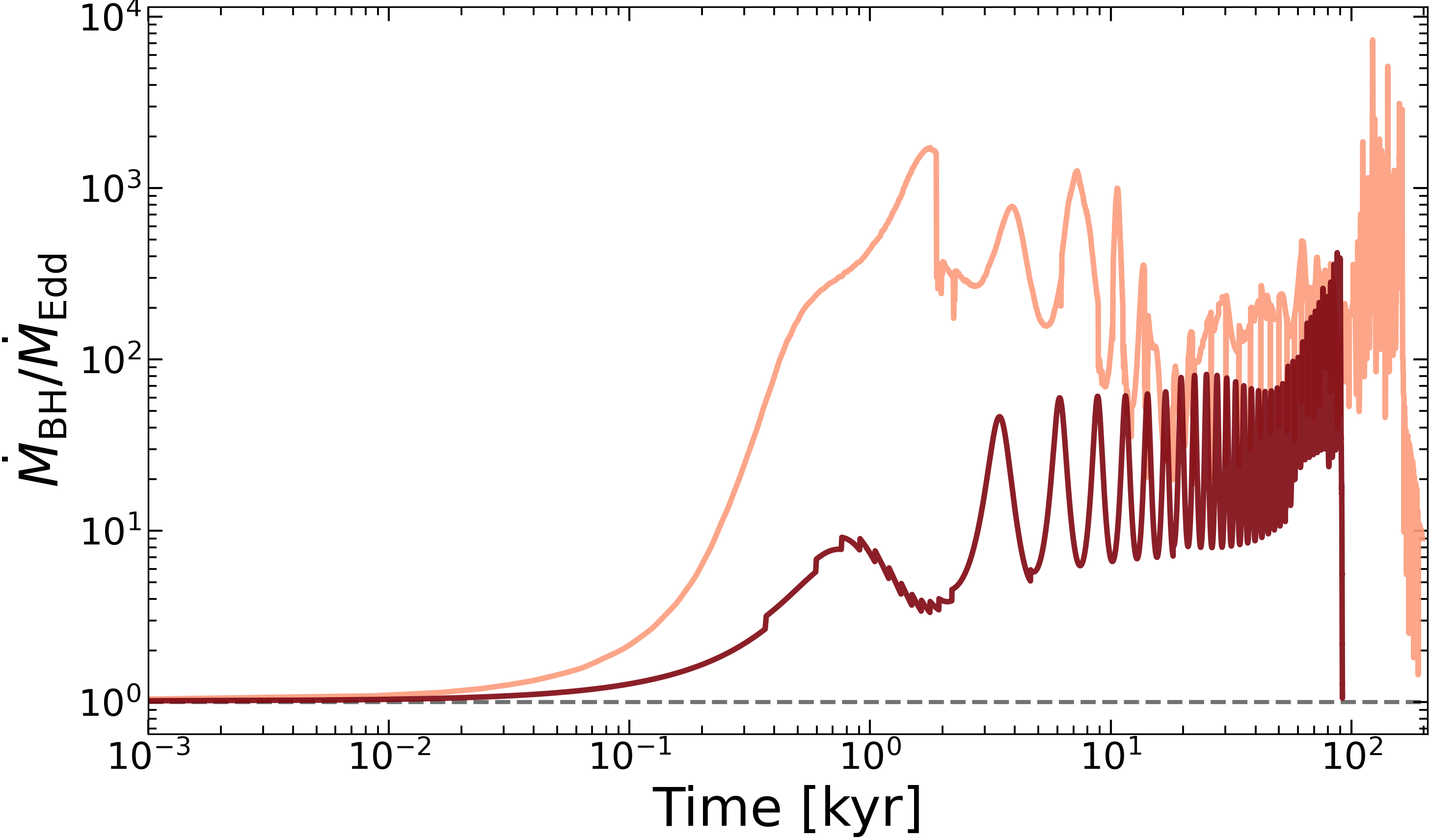}}
    \caption{Comparison of the initial SE episodes for the BHs in Run4. We time-shifted the onset of the SE accretion phase for the BH with initial mass $10^4$~M$_\odot$ and the first SE episode of the BH with initial mass $10^3$~M$_\odot$ to a common reference time. The SE phase in the former case reaches higher accretion rates and persists for a longer duration, accounting for the more rapid mass growth observed.}
    \label{fig:Run4Peaks}
\end{figure}

When no BH feedback is considered, we observe a significant growth in BH mass, particularly for the BH with an initial mass of $10^4$~M$_\odot$, which increases to $\sim 2\times10^5$~M$_\odot$ following a substantial SE accretion episode occurring at the time of seeding [see panels (b) and (d) in Figure~\ref{fig:feedback_comparison}, and also panel (d) of Figure~\ref{fig:all_runs}]. In the case of the lower-mass BH, we also observe multiple SE accretion episodes, leading to enhanced accretion compared to the run that includes feedback. However, in this case the final BH mass does not become significantly larger than the initial one. This difference can be understood in terms of the adopted BHL accretion model, in which the accretion rate scales as $M_{\rm BH}^2$ (see Equation~\ref{eq:bhl}). As a result, even SE phases of similar duration lead to significantly different growth histories. Indeed, as shown in Figure~\ref{fig:Run4Peaks}, the more massive BH consistently accretes at rates more than an order of magnitude higher than its lower-mass counterpart. In addition, given the relatively low masses and the turbulent nature of the host haloes at these redshifts, the gas supply becomes stochastic. For instance, the interplay between the BH and its host galaxy can be affected by slightly different seeding times or positions, as well as by the presence or absence of nearby SF, all of which can influence the overall BH accretion history. Also gas supply is further suppressed by stellar feedback at later times.

We argue that the feedback responsible for the limited accretion is primarily the kinetic component, as shown below. Nevertheless, from these first four runs it is already evident that the major merger occurring in all the cases analysed so far does not significantly affect the accretion rate. 
The only noticeable effect is a very brief accretion episode in Run4 reaching values close to the Eddington rate [see $z\sim11$, panel (d) in Figure~\ref{fig:seeding_comparison}]. The absence of BH feedback allows the more massive BH seed to sustain an accretion rate that increases with decreasing redshift, from $z\sim19$ to $z=10.3$. At this stage, the merger efficiently channels gas towards the central region of the halo, enhancing the gas density and driving the accretion rate to $\sim2$ times the Eddington limit. Shortly afterward, supernova (SN) feedback abruptly suppresses the accretion, causing the accretion rate to drop by several orders of magnitude.

\subsection{Extension to lower redshifts}\label{subsection:z8}

\begin{figure*}
    \centering
    \includegraphics[width=1.0\linewidth]{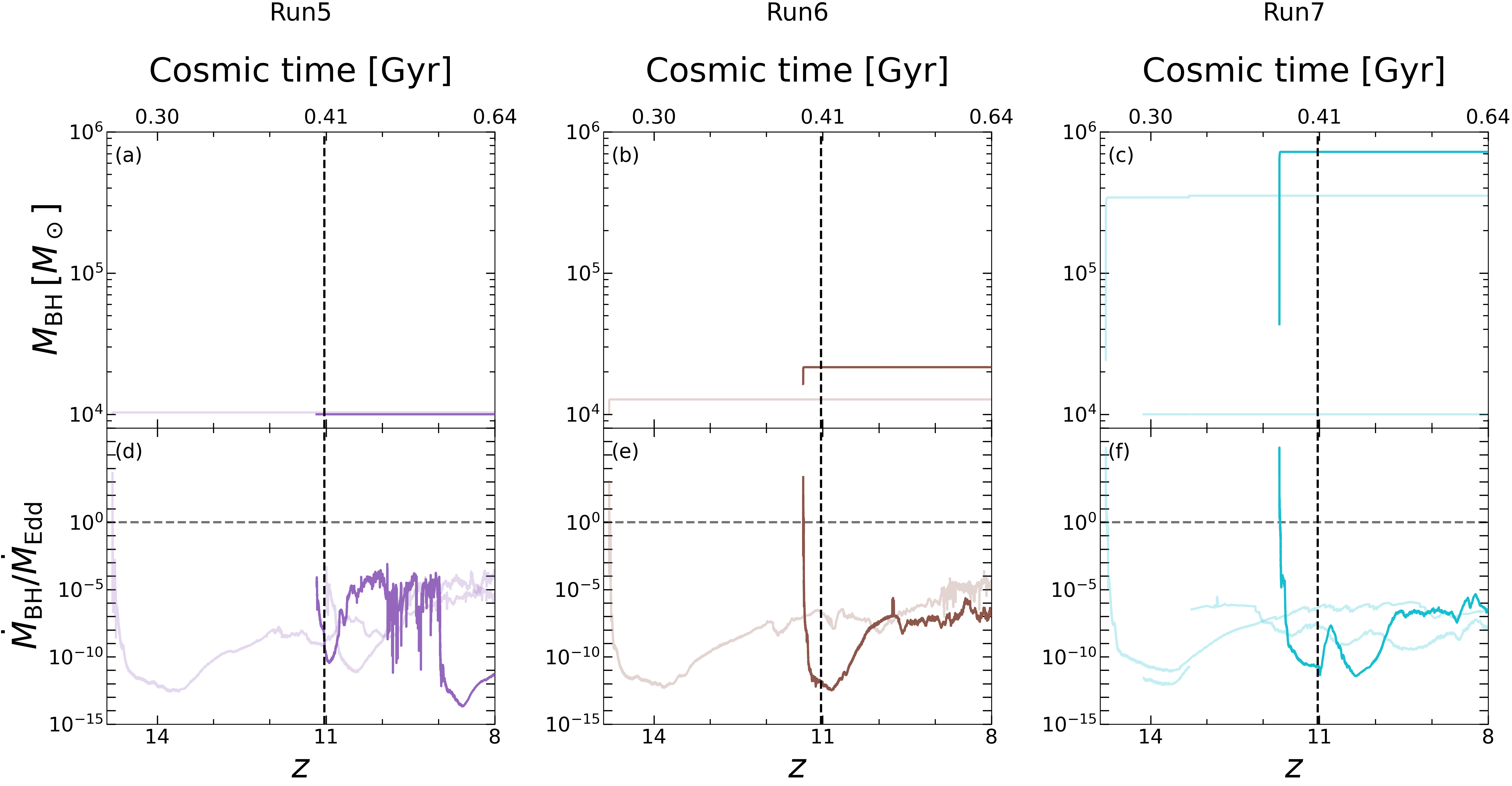}
    \caption{Comparison between the extended runs. We show, again, the evolution of BH masses and the BH accretion rates as in Figure~\ref{fig:seeding_comparison}. In the left-hand panels, we show the result with both kinetic and radiative feedback, with spin $a=0.7$. Black holes do not experience significant accretion, except for the first BH that has a brief burst of SE accretion, but the gained mass is negligible with respect to the initial one. Lowering the spin to $a=0.2$, and thus the feedback, enhances the accretion (central panels), but in the best case the final mass is approximately two times the initial one. Finally, in the right-hand panels, we show the case neglecting the kinetic feedback. We see here that two of the seeded BHs grow by a factor of 10--100, showing that the main reason of their suppressed growth in previous runs was the presence of the kinetic feedback. The short curve in the bottom-right panel terminates at $z \simeq 13.3$ because the corresponding BH merges with the more massive BH, whose mass at that time is $\sim 3 \times 10^{5} ~\mathrm{M}_{\odot}$. The merger produces the small simultaneous increase in the mass of the latter. However, we note that in no case did we see an increase in accretion rate due to the merger.}
    \label{fig:z8}
\end{figure*}

To further investigate the impact of mergers on BH accretion, we performed a set of simulations with $z_{\rm end}=8$ in order to verify whether the limited effect observed in the previous runs was due to the relatively early termination of the simulations. The results are shown in Figure~\ref{fig:z8}. We adopted the same seeding prescription as in Run1, but, motivated by the results of Run3, we seeded the BHs at the density peak of the haloes rather than at the potential minimum. In particular, placing the seeds at the density peaks is expected to maximise the initially available gas supply, thereby generally favouring the initial SE accretion phase, as visible in Figure~\ref{fig:z8} [see panels (d), (e), and (f)].

We note that, because BH formation is triggered when the host reaches the adopted stellar mass threshold, differences in the stellar mass growth histories also lead to different seeding times. In particular, the main halo BHs in these runs are seeded only at $z\sim11$--$12$, close to the merger redshift identified by the halo finder. However, this temporal proximity does not imply that the high initial accretion rates are already driven by the interaction between the two gaseous components. The merger time is defined from the total gravitational potential of the system, which is largely dominated by the DM component, whereas the two central gas reservoirs require an additional dynamical time\footnote{This is much larger than the typical accretion timescales we find in all of our simulations ($\sim$$10$--$100$~kyr).} to interact directly and mix. Shortly after BH seeding, the BH is still embedded in the compact gas clump of its original host, while the gas reservoir associated with the companion halo remains spatially distinct. Thus, the dense gas environment surrounding the newly seeded BH has not yet been enhanced by merger-driven inflows.

We performed three runs:

\begin{itemize}

    \item Run5, with the BH spin assumed as in the first three runs: $a = 0.7$;
    
    \item Run6, with a reduced BH spin of $a=0.2$;
    
    \item Run7, in which the kinetic feedback component was completely switched off.
    
\end{itemize}

Even when extending the simulation to lower redshift, Run5 does not show significant differences with respect to Run1 in terms of BH mass growth. Seeding at the density peak leads, in the case of the BH seeded at $z\sim15$, to an initial SE phase; however, this is not sufficient to significantly increase the BH mass.

\begin{figure*}
    \centering
    \includegraphics[width=17cm]{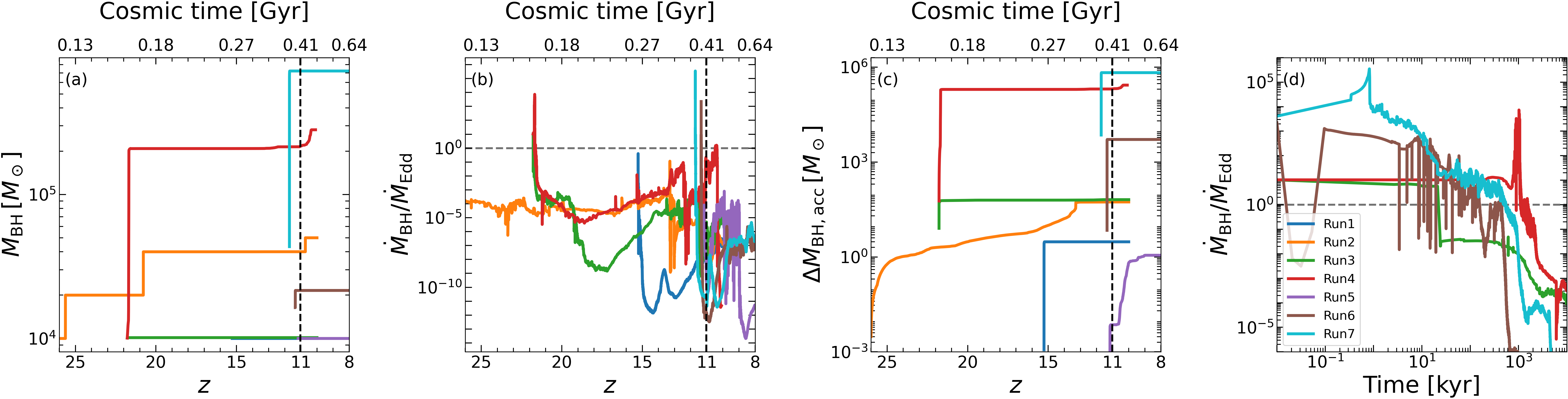}
    \caption{Evolution of the properties of the BH that is the most massive at the final available snapshot of each simulation and is then traced backwards throughout its history. Panel (a) shows its mass, whereas panel (b) shows its accretion rate in units of the Eddington rate. Panel (c) presents the cumulative BH mass gained through gas accretion alone, obtained by integrating $\dot{M}_{\rm BH}$ from the seeding time, this helps distinguish growth through gas accretion from mass increases caused by BH mergers. To facilitate a direct comparison of the SE phases, panel (d) shows the SE episodes of the same BHs, time-shifted to a common starting point, allowing their duration and amplitude to be compared.}
    \label{fig:all_runs}
\end{figure*}

In Run6, the lower spin value facilitates the initial accretion phase [panels (a) and (b) of Figure~\ref{fig:z8}]. However, the accretion is still sustained for just a $\sim$$10^2$~kyr [see panel (d) of Figure~\ref{fig:all_runs}], and even the most massive BH reaches a final mass only a few times larger than the initial seed mass.

As discussed earlier, feedback is the main mechanism regulating both the intensity and the duration of accretion, particularly during the initial SE phase. The results of Run7 further show that most of this effect is driven by the kinetic component: neglecting it leads to BHs that, after an initial SE phase, grow from an initial mass of $10^4$~M$_\odot$ to $10^{5-6}$~M$_\odot$. This is consistent with the behaviour observed in the run without kinetic feedback in isolated galaxies \citep[][]{Zana2025}.

Even in these extended simulations, mergers do not significantly enhance BH growth. The only exception remains Run4 [panel (d) of Figure~\ref{fig:feedback_comparison}], in which all forms of feedback are neglected. However, in this case SE accretion is stopped by SN explosions.
In order to compare the results for all the different runs, in Figure~\ref{fig:all_runs} we show the mass and accretion history over time for the most massive BH in each run, also comparing the intensity and time duration of the SE phases. 

\section{Discussion and conclusions}\label{section: discussion&conclusions}

We performed high-resolution cosmological zoom-in simulations of a major merger between two relatively small galaxies, compared to the quasar host investigated by \citet{Lupi2024a}, in order to test whether such events can trigger sustained SE accretion in massive BHs. In addition, we explored different BH seeding prescriptions and feedback models to assess their impact on the BH growth history.

The results presented in Section~\ref{subsection: Seeding dependence} show that the adopted seeding prescription plays an important role in triggering SE accretion at the moment of BH formation. In particular, seeding in dense gas environments can lead to short initial SE bursts. However, this initial phase alone is not sufficient to sustain high levels of accretion over long timescales. With the feedback prescriptions adopted in this work, strong kinetic feedback suppresses BH growth in the low-mass haloes considered here. Amongst the first three runs, only Run2 produces a final BH mass substantially larger than the initial seed mass, but this increase is driven mainly by hierarchical BH mergers rather than by gas accretion, as shown in panel (b) of Figure~\ref{fig:seeding_comparison}. This is consistent with results from the BRAHMA simulation suite \citep[][]{Bhowmick2024}, in which BH growth is dominated by mergers at high redshift.

The limited gas accretion in our case cannot be attributed solely to the lack of available gas due to stellar feedback. In both Run2 and Run3, BHs are seeded in haloes that do not yet contain stellar particles, which suggests that SN feedback does not play a dominant role at the moment of seeding. Although, in small haloes, a single SN explosion can completely disrupt the gas reservoir, the SE episodes occurring just after seeding observed in our simulations are relatively short ($\sim$$10^{1-3}$~kyr), and therefore require finely tuned timing to be systematically suppressed by SNe. Stellar feedback is instead more relevant at later times, contributing to the suppression of subsequent BH growth by keeping the accretion rate low. This is consistent with Run4, where SF is strongly suppressed (see Figure~\ref{fig:halo_properties}) and the accretion rate increases towards lower redshift, even reaching SE accretion rates, until a SN feedback event at $z\sim 10.3$ causes a sharp drop of several orders of magnitude.

The results presented in Sections~\ref{subsection: feedback_dependence} and \ref{subsection:z8} indicate that BH feedback is the primary mechanism regulating gas accretion onto BHs in these systems. In particular, the strength of the kinetic feedback plays a crucial role, consistent with the findings of \citet{Kung-Yi2023, Kung-Yi2025}. When the kinetic component is neglected [see panels (c) and (f) in Figure~\ref{fig:z8}], accretion rates increase and persist over longer timescales - $\sim$$10^3$~kyr, see panel (d) of Figure~\ref{fig:all_runs} - and BHs grow significantly, reaching masses of $10^{5-6}$~M$_\odot$. This behaviour is consistent with other cosmological simulations \citep[][]{Prole2026, Chon2026} and isolated galaxy simulations \citep[][]{Toyouchi2021,Zana2025}, in which feedback is purely radiative and accretion is modelled using the BHL prescription. In contrast, when kinetic feedback is included, the initial SE phase is rapidly quenched and BH growth remains strongly suppressed.

The behaviour observed in runs including kinetic feedback is directly linked to the adopted prescription. Although physically motivated, as it is based on the Blandford-Znajek mechanism \citep[][]{Blandford&Znajek1977}, it still presents some limitations. In particular, while we included a variable magnetic flux (Equation~\ref{eq.:magnetic flux}), the BH spin was kept fixed. This parameter is crucial for determining the efficiency of kinetic feedback (see Equation~\ref{eq.:kin eff}). A low-spin BH undergoing SE accretion may spin down through the extraction of rotational energy, potentially leading to an initial phase more similar to the case without kinetic feedback, or even to hyper-Eddington accretion \citep[][]{Inayoshi2016}.

A simple energetic argument further illustrates the strength of this feedback. The power of the jet can be written as

\begin{equation}
    P_{\rm jet}=\eta_{\rm jet}\dot{M}_{\rm BH}c^2.
\end{equation}

\noindent For very high accretion rates ($f_{\rm Edd}\gg f_{\rm c}$), the jet efficiency becomes primarily a function of the BH spin, $\eta_{\rm jet}(a)$ (see Equations~\ref{eq.:magnetic flux}, \ref{eq.:mad}, \ref{eq.:kin eff}, and \ref{eq.:ang_mom}). Assuming a short time interval $\Delta t$ and writing $\dot{M}_{\rm BH}=\lambda\dot{M}_{\rm Edd}$, the total injected energy becomes

\begin{equation}
E_{\rm jet}
\simeq
6.3\times10^{54}\,
\eta_{\rm jet}(a)\,
\left(\frac{M_{\rm BH}}{10^4\,{\rm M}_\odot}\right)
\left(\frac{\lambda}{10}\right)
\left(\frac{\Delta t}{\rm kyr}\right)
\ {\rm erg},
\end{equation}

\noindent where the spin dependence is enclosed in $\eta_{\rm jet}(a)$. For reference, the efficiencies corresponding to the spins adopted in our simulations are $\eta_{\rm jet}(0.7)\approx0.665$ and $\eta_{\rm jet}(0.2)\approx0.0175$. Also $\eta_{\rm jet}(0)=0$, as expected, since no rotational energy can be extracted from a non-spinning BH. This estimate is valid only for short, high-accretion episodes, since the BH spin, in a realistic environment, evolves due to angular momentum extraction.

Under these assumptions, the injected energy can easily exceed the binding energy of the host halo. For the haloes considered here, with $M_{\rm halo}\approx10^9$~M$_\odot$ and $R_{\rm halo}\approx5$ kpc at $z=10$, the binding energy is $E_{\rm bind}\sim10^{55}$ erg. An SE burst with $\lambda\sim10$--100, lasting even a few kilo years, can therefore inject energy comparable to the binding energy of the halo into the gas surrounding the BH, potentially ejecting the gas from the halo. 

In addition, several numerical effects may further suppress accretion in our simulations:

\begin{itemize}

\item The feedback particles are launched with high velocity ($v_{\rm jet}=0.1c$) and with a large mass-loading factor\footnote{If $f_{\rm Edd}\gg f_c$, then $\phi\sim\phi_{\rm MAD}$ (see Equations~\ref{eq.:magnetic flux}--\ref{eq.:mad}), so $\beta$ becomes a function of just the spin $a$; for $a=0.7$, $\beta\simeq138$, whereas for $a=0.2$, $\beta\simeq7$.}. As a result, the gas contained within the accretion kernel may be rapidly depleted.

\item Even when the kernel is not completely emptied, feedback reduces the gas density and increases its temperature, lowering the BHL accretion rate by several orders of magnitude (Equation~\ref{eq:bhl}).

\item A fraction of the accreted mass, determined by $\eta_{\rm jet}$, is converted into kinetic energy and distributed amongst neighbouring gas particles, effectively removing mass from the accretion flow.

\item Our resolution, although high, does not resolve the size of a jet at its base, and therefore a larger
volume is impacted \citep{Takeo2020}.

\end{itemize}

Another important result of our simulations is that the only significant SE accretion episodes occur immediately after BH seeding. Even when extending the simulations to $z=8$ to capture possible delayed effects of the merger, we do not find sustained SE accretion over long timescales. This differs from the behaviour found in more massive quasar-host environments \citet{Lupi2024a}. This difference is due to the lower halo masses considered here, which produce a much shallower gravitational potential well. In these environments, BH feedback suppresses gas inflow efficiently, and the initial SE burst appears to be the only viable route to substantial BH growth when kinetic feedback is absent or weak.

The merger itself does trigger gas inflows towards the central regions and increases the gas density, as indicated by the sharp rise in stellar mass at $z \sim 11$ in Figure~\ref{fig:halo_properties}. However, this is not sufficient to produce a sustained enhancement in the accretion rate, for two reasons. First, the timescale required to build up a high-density gas reservoir around the BH is not short, and in such low-mass haloes even a single SN explosion can disperse the gas before sustained accretion is established. This is clearly illustrated by Run4 (Figure~\ref{fig:feedback_comparison}), where the post-merger accretion rate rises up to the Eddington limit, but then drops by several orders of magnitude at $z \sim 10.3$ following SN explosions. Second, in the other runs, radiation pressure and kinetic feedback partially counteract the compression of the gas, preventing the central density around the BH from reaching the values required for strong accretion.

We argue that this behaviour is linked to the adopted accretion prescription (Equation~\ref{eq:bhl}), in which the accretion rate depends linearly on the local gas density and quadratically on the BH mass. In Run4, the high post-merger accretion rates are also partly due to the fact that the BH had already grown to $\sim$$10^5$~M$_\odot$, which naturally yields accretion rates about two orders of magnitude higher than those of BHs that remained near the initial seed mass of $10^4$~M$_\odot$. In this respect, a further difference with \citet{Lupi2024a} is that their simulations adopted an initial BH seed mass of $10^5$~M$_\odot$.

\begin{figure}
    \resizebox{\hsize}{!}{\includegraphics{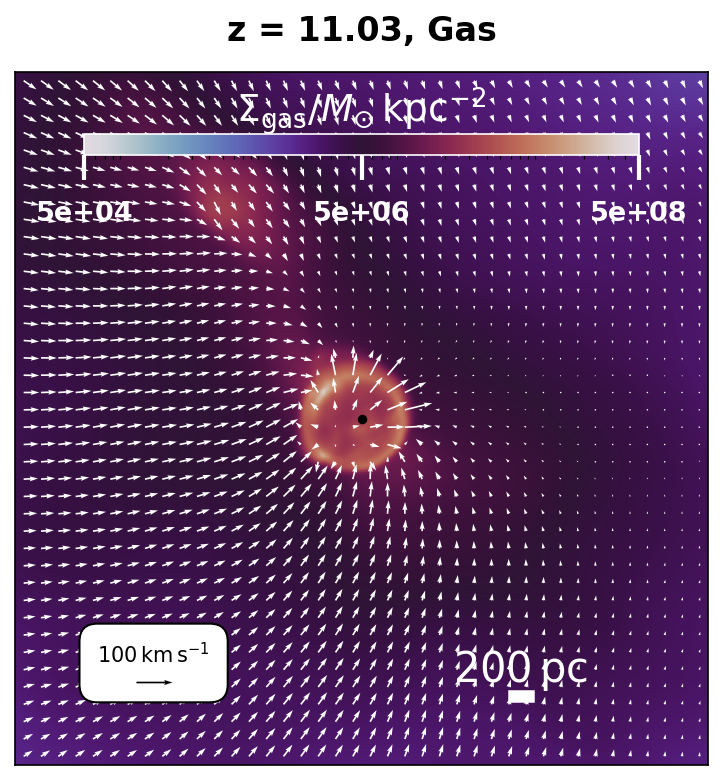}}

    \vspace{0.3cm}

    \resizebox{\hsize}{!}{\includegraphics{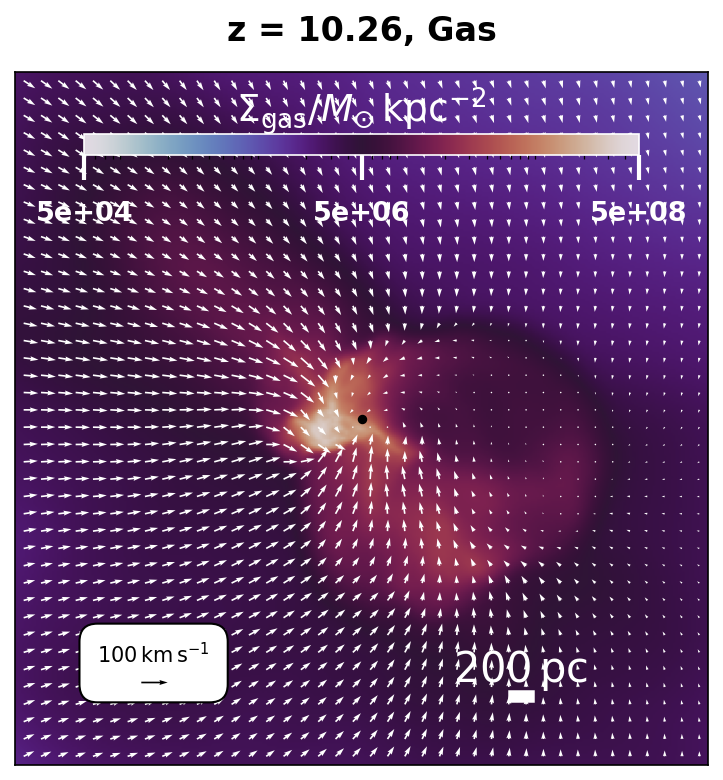}}

    \caption{Surface gas density maps of Run5 during the merger event, centred on the most massive BH (black dot), with the gas velocity field overlaid. The surface density and velocity field were computed by projecting gas within a sphere of radius $10\,{\rm kpc}$ centred on the BH, corresponding to a maximum line-of-sight depth of $20\,{\rm kpc}$. The velocity vectors show the gas velocity field in the simulation frame after recentring on the BH position. All lengths are given in physical units. In the top panel, shortly after BH seeding, a feedback-driven outflow expels gas from the central region. In the bottom panel, after the merger between the two haloes, gas inflows towards the centre are triggered, leading to the accumulation of gas around the BH.}
    
    \label{fig:merger time}
\end{figure}

We show in Figure~\ref{fig:merger time} two snapshots corresponding to key stages in the evolution of the most massive BH in Run5. The top panel shows the system immediately after the BH seeding event, but prior to the halo merger. At this stage, the feedback-driven outflow is clearly visible, as gas is expelled from the vicinity of the BH. The bottom panel shows a later stage, following the merger of the two  haloes. In this case, the merger triggers strong gas inflows towards the central region hosting the BH, leading to a significant accumulation of gas around it.

Finally, the high number density of massive BHs recently revealed by JWST suggests that a key ingredient is still missing. These BH masses are highly uncertain, and may be overestimated, as discussed in \citet{Lupi2024b, Trinca2026}. Another possibility is that something is still missing from the current theoretical picture. This may be related to either a better understanding of the BH spin distribution at seed formation and its subsequent evolution or the accretion prescriptions adopted in cosmological simulations. In this work, as in most numerical studies, we use the BHL prescription (Equation~\ref{eq:bhl}). However, this model assumes spherical symmetry, negligible angular momentum, and stationary gas assumptions that may not hold in the highly turbulent environments typical of low-mass haloes and may fail to capture the complex gas inflows triggered by major mergers. 
Furthermore, the BHL accretion rate scales quadratically with the BH mass, naturally favouring more massive seeds and thereby preventing sustained high accretion rates at high redshifts \citep{AnglesAlcazar2013, Hobbs2012}. Alternative accretion models that account for chaotic cold accretion, angular-momentum transport, or mass fluxes within the BH kernel \citep[][]{Gaspari2013, Zhu2022} may therefore provide a more realistic description of BH growth in these systems.

\begin{acknowledgements}
The authors are grateful to the anonymous referee for the
constructive comments and suggestions, which helped improve the quality of
the manuscript. We acknowledge ISCRA for awarding this project access to the LEONARDO supercomputer, owned by the EuroHPC Joint Undertaking, hosted by CINECA (Italy). RC acknowledges support with ``Progetti per Avvio alla Ricerca - Tipo 1'' from Sapienza University of Rome, grant AR125199BE8062A3. RC, RS and TZ acknowledge support from the MUR projects FIS-2024-01621 DAWN and PRIN 2022CB3PJ3-FLAGS, from EU-Recovery Fund PNRR, and from the INFN TEONGRAV initiative.
PRC acknowledges support from the Swiss National Science Foundation under the Sinergia Grant CRSII5\_213497 (GW-Learn). AT acknowledges financial support from the Bando Ricerca Fondamentale INAF 2023, Mini-grant ``Cosmic Archaeology with the first black hole seeds'' (Ob.Fu. RSN1 1.05.23.04.01).
\end{acknowledgements}

\bibliographystyle{aa} 
\bibliography{references}

\begin{appendix}

\section{Baryonic physics}\label{Appendix:baryonic_physics}

Our simulations were performed adopting physically motivated sub-grid prescriptions for baryonic physics, self consistently computing the chemistry of the most abundant species in the interstellar medium, SF, and stellar feedback in detail. In the following, we provide a brief description of the relevant physics, that is the same described in \citet{Lupi2024a}.

\subsection{Chemical network}

The chemical network regulating the heating and cooling processes was modelled using \textsc{krome} \citep[][]{KROME}. This includes non-equilibrium chemistry for nine primordial species (H, H$^{+}$, He, He$^{+}$, He$^{++}$, H$^{-}$, H$_2$, H$_2^{+}$, and e$^{-}$), with H$_2$ formation occurring both through H$^{-}$ associative detachment and on dust grains. Dust was not evolved as an independent component. Instead, the dust
abundance entering the chemistry and shielding prescriptions is inferred
from the local gas metallicity assuming a fixed dust-to-metal ratio. The network is further extended to include high-ionisation states of several important species commonly observed in quasar hosts, namely C[I-IV], O[I-VI], N[I-V], and Fe[I-II], also accounting for their contribution to the low-temperature cooling of the gas. Only oxygen and iron are explicitly produced and tracked by the stellar-enrichment model. The total metallicity used by the chemical network is derived from these tracked abundances, while the abundances of the other metal species, including carbon and nitrogen, are assigned assuming solar abundance ratios at the corresponding total metallicity. Their ionisation states are then evolved by the non-equilibrium chemical network. We additionally incorporate detailed X-ray chemistry in the network, accounting for the impact of AGN Compton heating, assuming $T_{\rm C,soft} = 3.23 \times 10^{6}\,\mathrm{K}$ and $T_{\rm C,hard} = 8.41 \times 10^{7}\,\mathrm{K}$ for soft and hard X-rays, respectively.

\subsection{Star formation}

A stochastic SF prescription was adopted, in which gas particles are probabilistically converted into stellar particles. The SFR density is defined as

\begin{equation}
\dot{\rho}_{\mathrm{SF}} = \epsilon \frac{\rho_{\rm g}}{t_{\mathrm{ff}}},
\end{equation}

\noindent where $\rho_{\rm g}$ is the local gas density, and

\begin{equation}
t_{\mathrm{ff}} = \sqrt{\frac{3\pi}{32 G \rho_{\rm g}}}
\end{equation}

\noindent is the free-fall time. As in \citet{Lupi&Bovino2020}, the star formation efficiency $\epsilon$ is given by \citep[][]{Padoan2012}

\begin{equation}
\epsilon = \epsilon_0 \exp\left(-1.6 \frac{t_{\mathrm{ff}}}{t_{\mathrm{dyn}}}\right),
\end{equation}

\noindent where $\epsilon_0 = 0.9$ is the local SF efficiency \citep[][]{Semenov2016}, and

\begin{equation}
t_{\mathrm{dyn}} = \frac{L}{2 \sigma_{\mathrm{eff}}}
\end{equation}

\noindent is the dynamical time of the star-forming cloud. $L \simeq 0.5h$ is the particle grid-equivalent size, where $h$ is the particle smoothing length. The effective velocity dispersion is defined as $\sigma_{\rm eff} \equiv \sqrt{\sigma_{\rm turb}^2 + c_{\rm s}^2}$, and represents the total support against gravitational collapse. Following \citet{Lupi2024a}, we define $\sigma_{\rm turb}=||\nabla \otimes \mathbf{v}||/5$, where $\nabla \otimes \mathbf{v}$ is the velocity-gradient tensor within the cloud, and $c_{\rm s}$ is the sound speed. The SF density threshold was set to $\rho_{\rm g} / m_{\mathrm{H}} = 1\ \mathrm{cm}^{-3}$, with $m_{\mathrm{H}}$ the mass of an hydrogen atom, simply to avoid computing the SF rate in regions in which SF will never occur.

\subsection{Supernovae and winds}

SN feedback was implemented using the mechanical prescription of
\citet{Lupi2019}, based on the model of \citet{Hopkins2018}. In this
framework, SN explosions inject energy and momentum into the surrounding gas.

The initial ejecta momentum is
\begin{equation}
p_0 = \sqrt{2 M_{\rm ej} E_{\rm SN}},
\end{equation}
where \(M_{\rm ej}\) is the ejecta mass and \(E_{\rm SN}\) is the SN
explosion energy. Following \citet{Lupi2019}, the momentum coupled to the gas is
\begin{equation}
p(M_{\rm swept}) =
p_0
\left(
\frac{\min\{M_{\rm swept},M_{\rm cool}\}}{M_{\rm ej}}
\right)^{1/2},
\end{equation}
where \(M_{\rm swept}\) is the swept-up gas mass and \(M_{\rm cool}\) is the
cooling mass computed from the cooling radius of \citet{Martizzi2015}. The
momentum is therefore capped once \(M_{\rm swept}=M_{\rm cool}\), giving the
terminal momentum
\begin{equation}
p_{\rm fin} =
\sqrt{2 E_{\rm SN} M_{\rm cool}}.
\end{equation}
Following \citet{Lupi_et_al2019}, we further multiply this terminal momentum by
a factor of 2 to make SN feedback more effective.

Stellar particles represent simple stellar populations with a Kroupa initial mass function \citep[IMF;][]{Kroupa2001}, releasing mass and energy via stellar winds and type-II/Ia SNae. Type-II SN and stellar wind rates are computed using \textsc{starburst99} \citep[][]{Leitherer1999}, assuming progenitor masses in the range 8--40~M$_\odot$, while more massive stars collapse directly into BHs.

SN explosions are treated as discrete events injecting \(10^{51}\,\mathrm{erg}\). For type-II SNae, we adopt IMF-averaged ejecta properties:

\begin{equation}
M_{\rm ej} = 10.61 \, \tilde{Z}^{-0.096} \, {\rm M}_\odot,
\end{equation}

\noindent with \(M_{\rm oxy} = 1.021\,{\rm M}_\odot\), \(M_{\rm iron} = 0.106\,{\rm M}_\odot\), \(\mathbf{\tilde{Z} = \max(Z/{\rm Z}_\odot, 0.01)}\) and total metal mass

\begin{equation}
M_{\rm Z} = 2.09\,M_{\rm oxy} + 1.06\,M_{\rm iron} = 2.246\,{\rm M}_\odot.
\end{equation}

Type-Ia SNae follow the delay-time distribution of \citet{Maoz2012} over 0.1--10~Gyr, with per-event injection \(E_{\rm SN} = 10^{51}\,\mathrm{erg}\), \(M_{\rm ej} = 1.4\,{\rm M}_\odot\), \(M_{\rm oxy} = 0.14\,{\rm M}_\odot\), \(M_{\rm iron} = 0.63\,M_\odot\), and \(M_Z = 0.9604\,{\rm M}_\odot\). The type-II SN rate is approximated as
\begin{equation}
\dot{N}_{\rm SN} = 6.8 \times 10^{-3}
\left(\frac{t_{\rm Myr}}{t_{\max,{\rm Myr}}}\right)^{-0.648}
\frac{1}{t_{\max,{\rm Myr}}}
\quad {\rm Myr^{-1}\,M_\odot^{-1}},
\end{equation}

\noindent for \(t_{\min,{\rm Myr}} < t_{\rm Myr} < t_{\max,{\rm Myr}}\), where \(t_{\rm Myr}\) is the time in units of Myr and \(t_{\min,{\rm Myr}} = 5.09\), \(t_{\max,{\rm Myr}} = 38.1\). The type-Ia SN rate is

\begin{equation}
\dot{N}_{\rm SNIa} = 2.82 \times 10^{-4} \, t_{\rm Myr}^{-1}
\quad {\rm Myr^{-1}\,M_\odot^{-1}}.
\end{equation}

Stellar winds were implemented by injecting mass and thermal energy into the surrounding gas, assuming continuous rates derived from \textsc{starburst99}. For old stellar populations (\(t_{\rm Myr} \ge 100\)), the wind energy is computed as \(\dot{E}_{\rm w} = \frac{1}{2}\dot{M}_{\rm w} v_{\rm eff}^2\), with \(v_{\rm eff} = 30\,{\rm km\,s^{-1}}\).

A metallicity dependence is included for young stars as

\begin{equation}
\dot{M}_{\rm w,Z} =
\begin{cases}
\dot{M}_{\rm w} \tilde{Z}^{0.8} & t_{\rm Myr} < 2.7 \\
\dot{M}_{\rm w} \min\{1,\tilde{Z}^{0.8}\} & 2.7 \le t_{\rm Myr} < t_{\max}
\end{cases}
\end{equation}

\noindent and

\begin{equation}
\dot{E}_{\rm w,Z} =
\begin{cases}
\dot{E}_{\rm w} \tilde{Z}^{1.06} & t_{\rm Myr} < 2.7 \\
\dot{E}_{\rm w} \min\{1,\tilde{Z}^{1.06}\} & 2.7 \le t_{\rm Myr} < t_{\max}
\end{cases}.
\end{equation}

A time-step limiter for stellar particles is adopted to properly resolve discrete SN events.
Newly synthesised metals from AGB stars are not included in our enrichment model. Low-mass stars return mass and the metals already present in the stellar particles at formation, but no additional AGB nucleosynthetic yields are injected into the gas.

\subsection{Radiation}

On-the-fly radiative transfer was implemented following \citet{Lupi2020}, evolving radiation energy density and flux under a local approximation for the Eddington tensor. This results in a hyperbolic system solved with a Godunov-type method.

Stellar radiation is injected into the nearest $N_{\rm ngb} \approx 64$ gas neighbours using a kernel-weighted scheme. To account for unresolved absorption and prevent artificial radiation leakage, the injected photon number in each cell is attenuated according to the local column density estimated over an effective distance defined as the maximum between the source kernel size and the source--cell separation. Unresolved radiation pressure is included by imparting momentum kicks to neighbouring gas elements proportional to the absorbed photon energy, ensuring consistent coupling between radiation and gas dynamics.

The coupling between radiation and chemistry is handled with \textsc{krome}, which computes photoionisation and photoheating rates from the local photon flux. Gas opacities are treated self-consistently and coupled to the radiation transfer solver. Dust shielding and H$_2$ self-shielding are also accounted for in both the chemistry and radiation transfer modules using an effective absorption length-scale based on the local Jeans length, capped at $40 \,\rm K$. We also adopt the on-the-spot approximation and neglect radiation re-emitted by the gas.

\end{appendix}

\end{document}